\catcode`\@=11					



\font\fiverm=cmr5				
\font\fivemi=cmmi5				
\font\fivesy=cmsy5				
\font\fivebf=cmbx5				

\skewchar\fivemi='177
\skewchar\fivesy='60


\font\sixrm=cmr6				
\font\sixi=cmmi6				
\font\sixsy=cmsy6				
\font\sixbf=cmbx6				

\skewchar\sixi='177
\skewchar\sixsy='60


\font\sevenrm=cmr7				
\font\seveni=cmmi7				
\font\sevensy=cmsy7				
\font\sevenit=cmti7				
\font\sevenbf=cmbx7				

\skewchar\seveni='177
\skewchar\sevensy='60


\font\eightrm=cmr8				
\font\eighti=cmmi8				
\font\eightsy=cmsy8				
\font\eightit=cmti8				
\font\eightbf=cmbx8				

\skewchar\eighti='177
\skewchar\eightsy='60


\font\ninei=cmmi9
\font\ninesy=cmsy9

\skewchar\ninei='177
\skewchar\ninesy='60


\font\tenrm=cmr10				
\font\teni=cmmi10				
\font\tensy=cmsy10				
\font\tenex=cmex10				
\font\tenit=cmti10				
\font\tensl=cmsl10				
\font\tenbf=cmbx10				
\font\tentt=cmtt10				
\font\tenss=cmss10				
\font\tensc=cmcsc10				
\font\tenbi=cmmib10				

\skewchar\teni='177
\skewchar\tenbi='177
\skewchar\tensy='60

\def\tenpoint{\ifmmode\err@badsizechange\else
	\textfont0=\tenrm \scriptfont0=\sevenrm \scriptscriptfont0=\fiverm
	\textfont1=\teni  \scriptfont1=\seveni  \scriptscriptfont1=\fivemi
	\textfont2=\tensy \scriptfont2=\sevensy \scriptscriptfont2=\fivesy
	\textfont3=\tenex \scriptfont3=\tenex   \scriptscriptfont3=\tenex
	\textfont4=\tenit \scriptfont4=\sevenit \scriptscriptfont4=\sevenit
	\textfont5=\tensl
	\textfont6=\tenbf \scriptfont6=\sevenbf \scriptscriptfont6=\fivebf
	\textfont7=\tentt
	\textfont8=\tenbi \scriptfont8=\seveni  \scriptscriptfont8=\fivemi
	\def\rm{\tenrm\fam=0 }%
	\def\it{\tenit\fam=4 }%
	\def\sl{\tensl\fam=5 }%
	\def\bf{\tenbf\fam=6 }%
	\def\tt{\tentt\fam=7 }%
	\def\ss{\tenss}%
	\def\sc{\tensc}%
	\def\bmit{\fam=8 }%
	\rm\setparameters\setbaselines\fi}


\font\twelverm=cmr12				
\font\twelvei=cmmi12				
\font\twelvesy=cmsy10	scaled\magstep1		
\font\twelveex=cmex10	scaled\magstep1		
\font\twelveit=cmti12				
\font\twelvesl=cmsl12				
\font\twelvebf=cmbx12				
\font\twelvett=cmtt12				
\font\twelvess=cmss12				
\font\twelvesc=cmcsc10	scaled\magstep1		
\font\twelvebi=cmmib10	scaled\magstep1		

\skewchar\twelvei='177
\skewchar\twelvebi='177
\skewchar\twelvesy='60

\def\twelvepoint{\ifmmode\err@badsizechange\else
	\textfont0=\twelverm \scriptfont0=\eightrm \scriptscriptfont0=\sixrm
	\textfont1=\twelvei  \scriptfont1=\eighti  \scriptscriptfont1=\sixi
	\textfont2=\twelvesy \scriptfont2=\eightsy \scriptscriptfont2=\sixsy
	\textfont3=\twelveex \scriptfont3=\tenex   \scriptscriptfont3=\tenex
	\textfont4=\twelveit \scriptfont4=\eightit \scriptscriptfont4=\sevenit
	\textfont5=\twelvesl
	\textfont6=\twelvebf \scriptfont6=\eightbf \scriptscriptfont6=\sixbf
	\textfont7=\twelvett
	\textfont8=\twelvebi \scriptfont8=\eighti  \scriptscriptfont8=\sixi
	\def\rm{\twelverm\fam=0 }%
	\def\it{\twelveit\fam=4 }%
	\def\sl{\twelvesl\fam=5 }%
	\def\bf{\twelvebf\fam=6 }%
	\def\tt{\twelvett\fam=7 }%
	\def\ss{\twelvess}%
	\def\sc{\twelvesc}%
	\def\bmit{\fam=8 }%
	\rm\setparameters\setbaselines\fi}


\font\fourteenrm=cmr12	scaled\magstep1		
\font\fourteeni=cmmi12	scaled\magstep1		
\font\fourteensy=cmsy10	scaled\magstep2		
\font\fourteenex=cmex10	scaled\magstep2		
\font\fourteenit=cmti12	scaled\magstep1		
\font\fourteensl=cmsl12	scaled\magstep1		
\font\fourteenbf=cmbx12	scaled\magstep1		
\font\fourteentt=cmtt12	scaled\magstep1		
\font\fourteenss=cmss12	scaled\magstep1		
\font\fourteensc=cmcsc10 scaled\magstep2	
\font\fourteenbi=cmmib10 scaled\magstep2	

\skewchar\fourteeni='177
\skewchar\fourteenbi='177
\skewchar\fourteensy='60

\def\fourteenpoint{\ifmmode\err@badsizechange\else
	\textfont0=\fourteenrm \scriptfont0=\tenrm \scriptscriptfont0=\sevenrm
	\textfont1=\fourteeni  \scriptfont1=\teni  \scriptscriptfont1=\seveni
	\textfont2=\fourteensy \scriptfont2=\tensy \scriptscriptfont2=\sevensy
	\textfont3=\fourteenex \scriptfont3=\tenex \scriptscriptfont3=\tenex
	\textfont4=\fourteenit \scriptfont4=\tenit \scriptscriptfont4=\sevenit
	\textfont5=\fourteensl
	\textfont6=\fourteenbf \scriptfont6=\tenbf \scriptscriptfont6=\sevenbf
	\textfont7=\fourteentt
	\textfont8=\fourteenbi \scriptfont8=\tenbi \scriptscriptfont8=\seveni
	\def\rm{\fourteenrm\fam=0 }%
	\def\it{\fourteenit\fam=4 }%
	\def\sl{\fourteensl\fam=5 }%
	\def\bf{\fourteenbf\fam=6 }%
	\def\tt{\fourteentt\fam=7}%
	\def\ss{\fourteenss}%
	\def\sc{\fourteensc}%
	\def\bmit{\fam=8 }%
	\rm\setparameters\setbaselines\fi}


\font\seventeenrm=cmr10 scaled\magstep3		


\newdimen\rp@
\newcount\@basestretchnum
\newskip\@baseskip
\newskip\headskip
\newskip\footskip


\def\setparameters{\rp@=.1em
	\headskip=24\rp@
	\footskip=\headskip
	\delimitershortfall=5\rp@
	\nulldelimiterspace=1.2\rp@
	\scriptspace=0.5\rp@
	\abovedisplayskip=10\rp@ plus3\rp@ minus5\rp@
	\belowdisplayskip=10\rp@ plus3\rp@ minus5\rp@
	\abovedisplayshortskip=5\rp@ plus2\rp@ minus4\rp@
	\belowdisplayshortskip=10\rp@ plus3\rp@ minus5\rp@
	\normallineskip=\rp@
	\lineskip=\normallineskip
	\normallineskiplimit=0pt
	\lineskiplimit=\normallineskiplimit
	\jot=3\rp@
	\setbox0=\hbox{\the\textfont3 B}\p@renwd=\wd0
	\skip\footins=12\rp@ plus3\rp@ minus3\rp@
	\skip\topins=0pt plus0pt minus0pt}


\def\setbaselines{\maxdepth=4\rp@\baselinestretch=\@basestretchnum}


\def\baselinestretch{\afterassignment\@basestretch\@basestretchnum}
\def\@basestretch{%
	\@baseskip=12\rp@ \divide\@baseskip by1000
	\normalbaselineskip=\@basestretchnum\@baseskip
	\baselineskip=\normalbaselineskip
	\bigskipamount=\the\baselineskip
		plus.25\baselineskip minus.25\baselineskip
	\medskipamount=.5\baselineskip
		plus.125\baselineskip minus.125\baselineskip
	\smallskipamount=.25\baselineskip
		plus.0625\baselineskip minus.0625\baselineskip
	\setbox\strutbox=\hbox{\vrule height.708\baselineskip
		depth.292\baselineskip width0pt }}



\def\makeheadline{\vbox to0pt{\baselinestretch=1000
	\vskip-\headskip \vskip1.5pt
	\line{\vbox to\ht\strutbox{}\the\headline}\vss}\nointerlineskip}

\def\makefootline{\baselineskip=\footskip\line{\the\footline}}

\def\big#1{{\hbox{$\left#1\vbox to8.5\rp@ {}\right.\n@space$}}}
\def\Big#1{{\hbox{$\left#1\vbox to11.5\rp@ {}\right.\n@space$}}}
\def\bigg#1{{\hbox{$\left#1\vbox to14.5\rp@ {}\right.\n@space$}}}
\def\Bigg#1{{\hbox{$\left#1\vbox to17.5\rp@ {}\right.\n@space$}}}


\mathchardef\alpha="710B
\mathchardef\beta="710C
\mathchardef\gamma="710D
\mathchardef\delta="710E
\mathchardef\epsilon="710F
\mathchardef\zeta="7110
\mathchardef\eta="7111
\mathchardef\theta="7112
\mathchardef\iota="7113
\mathchardef\kappa="7114
\mathchardef\lambda="7115
\mathchardef\mu="7116
\mathchardef\nu="7117
\mathchardef\xi="7118
\mathchardef\pi="7119
\mathchardef\rho="711A
\mathchardef\sigma="711B
\mathchardef\tau="711C
\mathchardef\upsilon="711D
\mathchardef\phi="711E
\mathchardef\chi="711F
\mathchardef\psi="7120
\mathchardef\omega="7121
\mathchardef\varepsilon="7122
\mathchardef\vartheta="7123
\mathchardef\varpi="7124
\mathchardef\varrho="7125
\mathchardef\varsigma="7126
\mathchardef\varphi="7127
\mathchardef\imath="717B
\mathchardef\jmath="717C
\mathchardef\ell="7160
\mathchardef\wp="717D
\mathchardef\partial="7140
\mathchardef\flat="715B
\mathchardef\natural="715C
\mathchardef\sharp="715D


\def\err@badsizechange{%
	\immediate\write16{--> Size change not allowed in math mode, ignored}}

\baselinestretch=1000
\tenpoint

\catcode`\@=12					
\catcode`\@=11
\expandafter\ifx\csname @iasmacros\endcsname\relax
	\global\let\@iasmacros=\par
\else	\immediate\write16{}
	\immediate\write16{Warning:}
	\immediate\write16{You have tried to input iasmacros more than once.}
	\immediate\write16{}
	\endinput
\fi
\catcode`\@=12


\def\rmb{\seventeenrm}

\def\singlespace{\baselineskip=\normalbaselineskip}
\def\halfspace{\baselineskip=1.5\normalbaselineskip}
\def\doublespace{\baselineskip=2\normalbaselineskip}


\def\AB{\bigskip\parindent=40pt
        \centerline{\bf ABSTRACT}\medskip\halfspace\narrower}
\def\AE{\bigskip\nonarrower\doublespace}
\def\nonarrower{\advance\leftskip by-\parindent
	\advance\rightskip by-\parindent}


\def\boxit#1{\vbox{\hrule\hbox{\vrule\kern3pt
	\vbox{\kern3pt#1\kern3pt}\kern3pt\vrule}\hrule}}

\def\hence{\leavevmode\hbox{\bf .\raise5.5pt\hbox{.}.} }

\def\dalemb#1#2{{\vbox{\hrule height.#2pt
	\hbox{\vrule width.#2pt height#1pt \kern#1pt \vrule width.#2pt}
	\hrule height.#2pt}}}
\def\gtorder{\mathrel{\raise.3ex\hbox{$>$}\mkern-14mu
             \lower0.6ex\hbox{$\sim$}}}
\def\ltorder{\mathrel{\raise.3ex\hbox{$<$}\mkern-14mu
             \lower0.6ex\hbox{$\sim$}}}

\newdimen\fullhsize
\newbox\leftcolumn
\def\twoup{\hoffset=-.5in \voffset=-.25in
  \hsize=4.75in \fullhsize=10in \vsize=6.9in
  \def\fullline{\hbox to\fullhsize}
  \let\lr=L
  \output={\if L\lr
        \global\setbox\leftcolumn=\columnbox\global\let\lr=R \advancepageno
      \else \doubleformat \global\let\lr=L\fi
    \ifnum\outputpenalty>-20000 \else\dosupereject\fi}
  \def\doubleformat{\shipout\vbox{
    \fullline{\box\leftcolumn\hfil\columnbox}\advancepageno}}
  \def\columnbox{\leftline{\vbox{\makeheadline\pagebody\makefootline}}}
  \tolerance=1000 }
\twelvepoint
\doublespace
{\nopagenumbers{
\rightline{~~~November, 2004}
\bigskip\bigskip
\centerline{\rmb Stochastic Collapse and Decoherence of a}
\centerline{\rmb Non-Dissipative Forced Harmonic Oscillator}
\medskip
\centerline{\it Stephen L. Adler
}
\centerline{\bf Institute for Advanced Study}
\centerline{\bf Princeton, NJ 08540}
\medskip
\bigskip\bigskip
\leftline{\it Send correspondence to:}
\medskip
{\singlespace\leftline{Stephen L. Adler}
\leftline{Institute for Advanced Study}
\leftline{Einstein Drive, Princeton, NJ 08540}
\leftline{Phone 609-734-8051; FAX 609-924-8399; email adler@ias.
edu}}
\bigskip\bigskip
}}
\vfill\eject
\pageno=2
\AB
Careful monitoring of harmonically bound (or as a limiting case, 
free) masses is the basis of current and future gravitational wave  
detectors, and of nanomechanical devices designed to access the quantum   
regime.  We analyze the effects of stochastic localization models 
for state vector reduction, and of related models for environmental 
decoherence, on such systems, focusing our analysis on the non-dissipative 
forced harmonic oscillator, and its free mass limit.    
We derive an explicit formula for the time evolution of the 
expectation of a general operator in the presence of stochastic reduction 
or environmentally induced decoherence, for both the non-dissipative  
harmonic oscillator and the free mass.  
In the case of the oscillator, we also give a formula 
for the time evolution of the matrix element of the stochastic 
expectation density matrix between general coherent states.  
We show that the stochastic expectation 
of the variance of a Hermitian operator in any unraveling of the stochastic 
process is bounded by the 
variance computed from the stochastic expectation of the density matrix, and 
we develop a formal perturbation theory for calculating expectation values of 
operators within any unraveling.  Applying our results to current 
gravitational wave interferometer detectors and nanomechanical systems, we 
conclude that the deviations from quantum mechanics predicted by the 
continuous spontaneous localization (CSL) model of state vector reduction   
are at least five orders of magnitude below the relevant standard quantum 
limits for these experiments.  The proposed LISA gravitational wave detector 
will be two orders of magnitude away from the capability of observing an 
effect.  

\AE
\bigskip\bigskip
\vfill\eject
\pageno=3
\centerline{{\bf 1.~~Introduction}}
\bigskip
Testing whether quantum mechanics is an exactly correct theory, or is an 
approximate theory from which there are small deviations, is a subject of 
current theoretical and experimental interest.  Significant bounds have 
been set [1] on deterministic, nonlinear modifications of the Schr\"odinger 
equation [2], and such modifications are also theoretically disfavored 
because they have been shown [3] to lead to the possibility of superluminal 
communication.  On the other hand, stochastic modifications to the 
Schr\"odinger equation have been extensively studied [4] as a way of 
resolving the measurement problem in quantum mechanics, and are known to 
be theoretically viable.  This raises the question of what bounds on the 
stochasticity parameters are set by current experiments, and what degree 
of refinement of current experiments will be needed to confront, and thus 
verify or falisfy, the stochastic models.  

The most extensively studied stochastic models are those based on the 
concept of localization [4,5], in which a stochastic, Brownian motion couples 
to the system center of mass degree of freedom.  Weak bounds on the 
stochasticity parameters for this type of model  
can already be set [6] from experiments [7] observing 
fullerene diffraction, and stronger (but far from definitive) bounds will  
be set [8] by a recently proposed experiment [9] that aims to coherently   
superimpose spatially displaced states of a small mirror attached to 
a cantilever.  Our aim in this paper is to analyze the effects of stochastic  
localization on another class of precision experiments, involving the careful 
monitoring of massive objects in gravitational wave detectors, and of   
microscopic oscillating beams in nanomechanical experiments.  To this 
end, we analyze the stochastic Schr\"odinger equation for 
a non-dissipative forced 
harmonic oscillator, focusing particular attention on the effects of the 
stochasticity terms on the quantum non-demolition variables of 
the oscillator. We also derive analogous formulas for the limiting case of  
a free mass, correcting a factor of 2 error in previous formulas given in 
the CSL literature.    Because the stochastic expectation of the density 
matrix in the localization model obeys a differential equation used as a 
model for environmental decoherence, our results are also relevant to the 
study of decoherence effects on both the forced oscillator and free 
mass systems.  Analyzing various experiments using our results, we
conclude that for the parameters of current  gravitational 
wave detectors and nanomechanical beams, only weak bounds will be 
set on the CSL model stochasticity parameters.  The proposed LISA 
gravitational wave detector should do better, but is still not expected to 
see an effect.  

This paper is organized as follows.  In Sec. 2 we give the basic stochastic 
Schr\"odinger equation to be analyzed, the corresponding 
pure state density matrix 
equation, and the simpler equation for the stochastic expectation of the 
density matrix (which is the usual mixed state density matrix).  The latter 
equation, we note, is also used as a model for environmental decoherence 
effects, and so its solution is of particular interest.  We also review 
briefly the basic ideas of quantum non-demolition measurements, leading to 
the identification of the non-demolition variables of the forced harmonic 
oscillator.  In Sec. 3 we give results for the time evolution of expectations 
of the non-demolition and other low order 
variables of the forced oscillator.  For 
comparison with the zero frequency limit of the oscillator, we give in 
Sec. 4 analogous results for a free mass, rederiving (and correcting) results 
already in the literature.  In Sec. 5 we 
give formulas for the time evolution of stochastic expectations of 
general operators for the forced non-dissipative oscillator 
and for its free mass limit, and additionally derive a formula for  
transition amplitudes of the oscillator, giving results that also apply 
to environmental decoherence effects. In Sec. 6 we 
consider stochastic fluctuations, and show that expectations of variances 
of observables can be bounded using our earlier calculations proceeding 
from the expectation of the density matrix.   
In Sec. 7 we set up a formal perturbative 
procedure for calculating stochastic fluctuation effects, and use the leading     
order results to interpret the inequality derived in Sec. 6.  Finally, in 
Sec. 8, we apply our results to make estimates for the effects of CSL models 
in gravitational wave detection and nanomechanical 
resonator experiments.  In Appendix A we review some It\^o calculus 
formulas, and in Appendix B we relate the formalism used in the text to 
the Lindblad density matrix evolution equation.  
\bigskip
\centerline{{\bf 2.~~Basic formalism: one dimensional oscillator}}
\bigskip
We start our analysis by considering a massive harmonic oscillator in 
one dimension, which in the three-dimensional case will describe the 
dynamics of one center-of-mass degree of freedom.  The oscillator Hamiltonian
is taken as 
$$H=\hbar \omega (a^{\dagger} a+{1\over 2}) + d(t)a^{\dagger} 
+\overline{d}(t) a~~~,\eqno(1)$$
with $\omega$ the oscillator angular frequency, $d(t)$ a complex  
$c$-number driving term, and $a,a^{\dagger}$  annihilation and creation 
operators obeying $[a,a^{\dagger}]=1$.  These operators are related to 
the oscillator mass $m$, coordinate $q$, and momentum $p$, by 
$$\eqalign{
a=&(m\omega/2\hbar)^{1\over 2}(q+ip/m\omega)~~~,\cr   
a^{\dagger}=& (m\omega/2\hbar)^{1\over 2}(q-ip/m\omega)~~~,\cr   
q=&\sigma (a +a^{\dagger})~~~,~~\sigma= (\hbar/2 m \omega)^{1\over 2}~~~,\cr
}\eqno(2a)$$ 
and the number of quanta $N$ in the oscillator is given by 
$$N=a^{\dagger}a~~~.\eqno(2b)$$
Discussions of quantum non-demolition experiments involving oscillators [10] 
also introduce the quantities 
$$\eqalign{
X_1=&q \cos \omega t - (p/m \omega) \sin \omega t~~~, \cr 
X_2=&q \sin \omega t + (p/m \omega) \cos \omega t~~~, \cr 
}\eqno(3a)$$
from which one easily finds 
$$\eqalign{
q+ip/m\omega=& (X_1+iX_2)e^{-i\omega t}~~~,\cr    
q-ip/m\omega=& (X_1-iX_2)e^{i\omega t}~~~,\cr    
X_1=&\sigma(ae^{i\omega t}+a^{\dagger} e^{-i\omega t})~~~,\cr  
X_2=&-i\sigma(ae^{i\omega t}-a^{\dagger} e^{-i\omega t})~~~.\cr  
}\eqno(3b)$$
Hence $X_{1,2}$ are quantum mechanical analogs of the classical amplitude 
of the oscillator, and when the external driving term $d(t)$ is zero they 
are conserved, as is the occupation number $N$.  Because these quantities 
are constants of the motion in the absence of external forces, 
measurements of them, while introducing 
uncertainties into the conjugate variables (which are the phase $\phi$ in 
the case 
of $N$, $X_2$ in the case of $X_1$, and $X_1$ in the case of $X_2$),  
do not feed the uncertainties in the conjugate variables 
back into the time evolution of the measured 
variable.  Hence the variables $N$, $X_1$, and $X_2$ can in principle be 
measured to an accuracy not limited by the uncertainty principle, and are 
called ``quantum non-demolition'' variables. 

Letting $|\psi_t\rangle$ be the oscillator wave function at time $t$, the 
standard Schr\"odinger equation is 
$$d|\psi_t\rangle= -(i/\hbar) H dt |\psi_t \rangle~~~.\eqno(4a)$$
We shall be interested in this paper in a 
class of models [11] for state vector 
reduction, which modify Eq.~(4a) by adding stochastic terms to the 
Schr\"odinger equation.  Specifically, we shall consider the evolution  
equation 
$$
d\,|\psi_{t}\rangle \; = \; \left[ -{i \over \hbar}\, H\, dt +
\sqrt{\eta}\, (q - \langle q \rangle)\, dW_{t} -
{\eta\over 2}\, (q - \langle q \rangle)^2 dt \right]
|\psi_{t}\rangle~~~,\eqno(4b)$$
where $H$ is given by Eq.~(1), and $\langle q \rangle
\equiv \langle \psi_{t} | q | \psi_{t} \rangle$ is the quantum mechanical
expectation of the position operator $q$ of the oscillator. Introducing  
the pure state density matrix $\hat \rho(t)=|\psi_t\rangle \langle \psi_t|$, 
we can also write $\langle q \rangle = {\rm Tr} q \hat \rho(t)$.   
The stochastic dynamics is governed by a
standard Wiener processes $W_{t}$, defined on a probability space
$(\Omega,
{\cal F}, {\bf P})$.  Using the rules of the It\^o calculus 
(see  Appendix A), 
the density matrix evolution corresponding to Eq.~(4b) is 
$$d\hat\rho=-{i\over \hbar} [H,\hat\rho] dt -{1\over 2} \eta  
[q,[q,\hat\rho]] dt+ \sqrt{\eta} [\hat \rho,[\hat \rho,q]] dW_t ~~~.
\eqno(5a)$$
Since this evolution equation obeys $\{d\hat\rho,\hat \rho\}+(d\hat\rho)^2
=d\hat\rho$, it preserves the pure state condition $\hat \rho^2=\hat \rho$.  
When statistics are accumulated by averaging many runs of an experiment, the 
relevant 
density matrix in the stochastic case is the ensemble expectation 
$\rho=E[\hat \rho]$, giving the mixed state density matrix 
which obeys the ordinary differential equation 
$$\eqalign{
{d\rho\over dt}=&-{i\over \hbar} [H,\rho] -{1\over 2} \eta  
[q,[q,\rho]] \cr
=& -{i\over \hbar} [H,\rho] -{1\over 2} \eta \sigma^2 
[a+a^{\dagger},[a+a^{\dagger},\rho]]~~~.\cr
}\eqno(5b)$$
This equation is of particular interest because (with a different value 
of the parameter $\eta$) it is also used [12] as a simple model for 
environmental decoherence effects.  
The calculations of this paper focus on analyzing Eqs.~(5a) and (5b) for the 
forced harmonic oscillator Hamiltonian of Eq.~(1).  
\bigskip
\centerline{\bf 3.~~ Stochastic expectations of oscillator observables}
\bigskip
We begin by considering the evolution equation of Eq.~(5b) for the mixed  
state density matrix $\rho$.  Letting $B$ be any time-independent 
operator, let us denote by $\langle\langle B \rangle\rangle$ the 
expectation computed 
with the mixed state density matrix,  
$$\langle\langle B \rangle\rangle = {\rm Tr} \rho B~~~.\eqno(6a)$$
{}For the time evolution of this expectation, we then find
$$\eqalign{
{d \langle\langle B \rangle\rangle \over dt} =& {\rm Tr} {d \rho \over dt} B    \cr
=& {\rm Tr} B 
\Big(-{i\over \hbar} [H,\rho] -{1\over 2} \eta  
[q,[q,\rho]]\Big) \cr
=&{\rm Tr} \Big( -{i\over \hbar} [B,H] -{1\over 2} \eta \sigma^2 
[[B,a+a^{\dagger}],a+a^{\dagger}] \Big) \rho ~~~,\cr
}\eqno(6b)$$
where we have made repeated use of cyclic permutation under the trace.  
Let us now make successively the choices $B=a,a^{\dagger},aa,a^{\dagger}
a^{\dagger},a^{\dagger}a,aa^{\dagger}=1+a^{\dagger}a$, corresponding to 
all quantities linear and quadratic in the creation and 
annihilation operators.
Then evaluating the single and double commutators in the final line of  
Eq.~(6b), a simple calculation gives for the two linear operators, 
$$\eqalign{
{d \over dt} {\rm Tr} \rho a =& -i \omega {\rm Tr} \rho a 
-{i \over \hbar}d(t)~~~,\cr
{d \over dt} {\rm Tr} \rho a^{\dagger} =& i \omega {\rm Tr} \rho a^{\dagger} 
+{i \over \hbar}{\overline d}(t)~~~,\cr
}\eqno(7a)$$
and for the four quadratic operators
$$\eqalign{
{d \over dt} {\rm Tr} \rho aa =& -2i\omega {\rm Tr} \rho aa 
-2{i \over \hbar} d(t) {\rm Tr} \rho a - \eta \sigma^2~~~,\cr
{d \over dt} {\rm Tr} \rho a^{\dagger}a^{\dagger} =& 2i\omega 
{\rm Tr} \rho a^{\dagger}a^{\dagger}  
+2{i \over \hbar} {\overline d}(t) {\rm Tr} \rho a^{\dagger} 
- \eta \sigma^2~~~,\cr
{d \over dt} {\rm Tr} \rho a^{\dagger}a =& -{i\over \hbar} d(t) 
{\rm Tr} \rho a^{\dagger} + {i \over \hbar} \overline{d}(t) {\rm Tr} \rho a 
+ \eta \sigma^2~~~,\cr
{d \over dt} {\rm Tr} \rho a a^{\dagger} =
& {d \over dt} {\rm Tr} \rho a^{\dagger}a ~~~.\cr
}\eqno(7b)$$

These equations can be immediately integrated to give 
$$\eqalign{
{\rm Tr} \rho(t) a =& e^{-i\omega t} \Big[ {\rm Tr} \rho(0)a 
-{i\over \hbar} \int_0^t du d(u) e^{i\omega u}\Big] ~~~,\cr
{\rm Tr} \rho(t) a^{\dagger} =& e^{i\omega t} \Big[ {\rm Tr} 
\rho(0)a^{\dagger} 
+{i\over \hbar} \int_0^t du \overline{d}(u) e^{-i\omega u}\Big] ~~~\cr
}\eqno(8a)$$ 
for the linear operators, and  
$$\eqalign{
{\rm Tr} \rho(t) aa=& e^{-2i\omega t} \Big[ {\rm Tr}\rho(0) aa
-\int_0^t dv e^{2i\omega v} \Big( 2{i\over \hbar} d(v) {\rm Tr} \rho(v) a 
+ \eta \sigma^2\Big) \Big]~~~, \cr
{\rm Tr} \rho(t) a^{\dagger}a^{\dagger}=& e^{2i\omega t} 
\Big[ {\rm Tr}\rho(0) a^{\dagger}a^{\dagger} 
+\int_0^t dv e^{-2i\omega v} \Big( 2{i\over \hbar} \overline{d}(v) 
{\rm Tr} \rho(v) a^{\dagger}  - \eta \sigma^2\Big) \Big]~~~,  \cr
{\rm Tr} \rho(t) a^{\dagger}a =& {\rm Tr} \rho(0) a^{\dagger} a 
-{i\over \hbar} \int_0^t dv \Big[ d(v) {\rm Tr} \rho(v) a^{\dagger} 
-\overline{d}(v) {\rm Tr} \rho(v) a \Big] + \eta \sigma^2 t~~~, \cr 
{\rm Tr} \rho(t) a a^{\dagger}=& 1+{\rm Tr} \rho(t) a^{\dagger} a~~~\cr 
}\eqno(8b)$$
for the quadratic operators.
We see that by substituting Eq.~(8a) into Eq.~(8b), we can reduce the 
expressions for the quadratic operators to quadratures.  (Proceeding in a  
similar fashion, it is easy to see that given any 
polynomial $P(a,a^{\dagger})$ of finite degree in the creation and 
annihilation operators, the expectation ${\rm Tr} \rho(t) P$ can be reduced 
to quadratures; for an explicit formula constructed by generating function 
methods, see Sec. 5.)  

Rather than exhibiting the full expressions for the 
expectations of the quadratic operators, we note that what we are 
most interested in is calculating the change in these quantities, denoted 
by $\delta$, arising from the ``decoherence'' 
term with coefficient $\eta$ in Eq.~(5b).  
{}From the fact that Eq.~(8a) contains no terms proportional to $\eta$, we 
see that there are no stochastic (or decoherence) 
effects on the linear operators,  
$$\eqalign{
\delta {\rm Tr}\rho(t) a =&0~~~,\cr
\delta {\rm Tr}\rho(t) a^{\dagger}=&0~~~,\cr 
}\eqno(9a)$$ 
while the effect of the $\eta$ term in Eq.~(5b) on the quadratic operators 
is simply given by 
$$\eqalign{
\delta {\rm Tr} \rho(t) aa =& -{\eta \sigma^2 \over \omega} e^{-i\omega t} 
\sin \omega t~~~,\cr
\delta {\rm Tr} \rho(t) a^{\dagger}a^{\dagger} =& 
-{\eta \sigma^2 \over \omega} e^{i\omega t} \sin \omega t~~~, \cr
\delta {\rm Tr} \rho(t) a^{\dagger} a=&\delta {\rm Tr} \rho(t) a a^{\dagger} 
= \eta \sigma^2 t~~~.\cr
}\eqno(9b)$$
Using the definitions of $X_{1,2}$ given in Eq.~(3a,b), we correspondingly 
find that 
$$\eqalign{
\delta {\rm Tr} \rho(t) X_1=&\delta {\rm Tr} \rho(t) X_2 =0~~~,\cr
\delta {\rm Tr} \rho(t) X_1^2=& 2\eta \sigma^4\Big(t-{\sin 2\omega t\over 
2 \omega} \Big)~~~,\cr
\delta {\rm Tr} \rho(t) X_2^2=& 2\eta \sigma^4\Big(t+{\sin 2\omega t\over 
2 \omega} \Big)~~~,\cr
\delta {\rm Tr} \rho(t)(X_1X_2+X_2X_1)=&-{4 \eta \sigma^4 \over \omega} 
\sin^2 \omega t~~~,\cr
\delta {\rm Tr} \rho(t) [X_1,X_2]=&0~~~.\cr
}\eqno(10)$$
We note that these formulas are exact (not just approximations to first order 
in $\eta$), since for all the operators $B$ considered above, we have 
$${\rm Tr}\rho(t) B={\rm Tr}\rho(t) B|_{\eta =0} + \delta {\rm Tr} \rho(t) B
~~~.\eqno(11)$$
\bigskip
\centerline{\bf 4.~~The free mass limit}
\bigskip
According to Eq.~(9b), the oscillator occupation number $N=a^{\dagger}a$ 
contains a term that grows linearly in time as $\eta \sigma^2 t$.  Since 
the occupation number contribution to the oscillator energy of Eq.~(1) 
is $\hbar \omega N$, and since $\sigma^2=\hbar/2 m \omega$ from Eq.~(2a),  
the oscillator energy contains a term that grows linearly in time as 
$$\delta E = \delta {\rm Tr} \rho(t) H = {\eta \hbar^2 t \over 2 m}~~~.
\eqno(12)$$
Because this formula is independent of the oscillator frequency $\omega$, it 
should also correspond to the energy increase of an 
unbound mass $m$ arising from 
the $\eta $ term in Eq.~(5b).  This can be calculated directly as follows. 
{}For an unbound mass in one dimension, the Hamiltonian is $H=p^2/2m$, 
and the density matrix evolution is given by the first line of Eq.~(5b).  
So we have, by the same reasoning that led to Eq.~(6b),  
$$\eqalign{
{d \over dt} \delta {\rm Tr} \rho H  =& 
-{\rm Tr} {1\over 4m} \eta  [[p^2,q],q] \rho   \cr
=&-{\rm Tr} {1 \over 4m} \eta  [-2i \hbar p,q] \rho  = \eta\hbar^2/2m~~~,\cr
}\eqno(13a)$$
giving 
$$\delta {\rm Tr} \rho H = \eta \hbar^2 t/2m~~~,\eqno(13b)$$ 
in agreement with the result calculated for the oscillator.  This result 
is a factor of two larger than the one quoted in the CSL 
literature [13]; for instance, Taylor expansion of Eq.~(3.36) of Ghirardi,   
Pearle, and Rimini (GPR) shows that for a uniform cube, 
their $\gamma \delta_i$ is 
the same as the parameter $\eta$ used here, and so their formula of 
Eq.~(3.38c), which states that ${d \over dt} \langle\langle P_i^2
\rangle\rangle 
= {1\over 2} \gamma \delta_i \hbar^2$ would correspond to ${d \over dt} 
\langle\langle H \rangle\rangle = \eta \hbar^2/4m$, in 
disagreement with our result of 
Eq.~(13a) and with the oscillator calculation of the preceding section.  
This error propagates through to Eqs.~(3.41a) through (3.41c) of GPR, 
all of which are a factor of 2 too small.  Thus, in our 
notation, their results should read 
$$\eqalign{
\delta {\rm Tr} \rho p^2 =& \eta \hbar^2 t~~~,\cr
\delta {\rm Tr} \rho (pq+qp)=& \eta \hbar^2 t^2/m~~~,\cr
\delta {\rm Tr} \rho q^2 =& \eta \hbar^2 t^3/(3 m^2)~~~.\cr
}\eqno(13c)$$
A rederivation of the second and third lines of Eq.~(13c) will be given in 
the next section.  
[These equations were first given, with a different identification 
of the proportionality constant $\eta$, in the GRW model [11].  
Philip Pearle has rechecked the calculations in the 
paper of GPR, and finds that a factor of 2 error was made in going 
from their Eq.~(3.36) to their Eq.~(3.38c); when corrected, their equations 
agree with our results of Eq.~(13c) above.]  
\vfill\eject
\bigskip
\centerline{\bf 5.~~Exact general formulas for 
decoherence effects on a forced}
\centerline{\bf  non-dissipative harmonic oscillator and on a free mass}

\bigskip
We have seen in Eqs.~(8a) and (8b) that the double expectations 
\big(stochastic 
and quantum, as defined in Eq.~(6a)\big) of low order polynomials in the 
oscillator creation and annihilation operators can be reduced to quadratures.  
To show 
that this is a general result, let us consider the generating function 
$$\eqalign{
K_{\alpha\beta}(t)=&{\rm Tr}\big( \exp(\alpha a^{\dagger}e^{-i\omega t}) 
\exp(\beta a e^{i\omega t}) \rho(t)\big) \cr
=&\sum_{n=0}^{\infty}\sum_{m=0}^{\infty} {\alpha^n \over n!}
  {\beta^m \over m!} 
e^{i \omega t (m-n)}  {\rm Tr} (a^{\dagger})^n a^m \rho(t)~~~,\cr
}\eqno(14a)$$
from which one can extract the expectations of arbitrary normal ordered 
operators formed from $a$ and $a^{\dagger}$. To proceed, we shall need 
the generalization of Eq.~(6b) to 
the case when the operator $B$ has an explicit time dependence, which  
reads 
$$\eqalign{
{d \langle\langle B \rangle\rangle \over dt} =& 
{\rm Tr}\big( {d \rho \over dt} B+\rho {\partial B \over \partial t} \big)\cr
=&{\rm Tr}\Big( {\partial B \over \partial t}  
-{i\over \hbar} [B,H] -{1\over 2} \eta \sigma^2 
 [[B,a+a^{\dagger}],a+a^{\dagger}] \Big)  \rho~~~.\cr
}\eqno(14b)$$
Applying this formula to Eq.~(14a), with 
$B=  \exp(\alpha a^{\dagger}e^{-i\omega t})
 \exp(\beta a e^{i\omega t}) $, 
the explicit time derivative on the 
right cancels the commutator term involving the free Hamiltonian $\hbar 
\omega a^{\dagger}a$ (this is why we included an explicit time dependence  
in the definition of the generating function), leaving the simple 
differential equation  
$${d \over d t} K_{\alpha \beta}(t) =
\left[ {i\over \hbar}\big(\alpha e^{-i\omega t} \overline d(t) 
- \beta e^{i \omega t} d(t) \big)  -{1\over 2}  \eta \sigma^2 
(\alpha e^{-i\omega t} - \beta e^{i\omega t})^2  \right] K_{\alpha \beta}(t)
~~~.\eqno(15)$$
Defining 
$$D(t)\equiv \int_0^t du e^{i\omega u}d(u)~~,~~~
\overline D(t) \equiv \int_0^t du e^{-i\omega u} 
\overline d(u)~~~,\eqno(16a)$$
the integral of Eq.~(15) takes the form
$$K_{\alpha \beta}(t) = \exp\left[ \alpha \beta \eta \sigma^2 t  
-{\eta \sigma^2 \over 2 \omega}
(\alpha^2 e^{-i \omega t} +\beta^2 e^{i \omega t}) \sin \omega t 
+{i \over \hbar}\big( \alpha \overline D(t) - \beta D(t) \big) \right] 
K_{\alpha \beta}(0)~~~,\eqno(16b)$$ 
with 
$$K_{\alpha \beta}(0)= {\rm Tr}\Big( e^{\alpha a^{\dagger}} e^{\beta a} 
\rho(0) \Big)~~~.\eqno(16c)$$ 
Expanding this equation through second order in $\alpha$ and $\beta$, one 
can verify that it agrees with the formulas of Eqs.~(8a) and (8b), and 
so we have obtained the generalization of these expressions to arbitrary 
normal ordered monomials in the creation and annihilation operators.  
Thus expectations of operators with respect to the 
density matrix of the decoherent forced 
oscillator can be explicitly calculated in closed form.  As an example of 
particular interest, we note that the $\eta$-dependent terms with the 
dominant time dependence for large times can be read off from the power 
series expansion of the first factor on the right hand side of Eq.~(16b), 
$$\exp(\alpha \beta \eta \sigma^2 t) = \sum_{n=0}^{\infty}   
{(\alpha \beta \eta \sigma^2 t)^n \over n!}~~~.\eqno(16d)$$
Thus, the leading $\eta$ dependence in ${\rm Tr} \rho(t) a^{\dagger}a$ 
at large times
is $\eta \sigma^2 t$, in agreement with Eq.~(8b), while the leading 
$\eta$ dependence in ${\rm Tr}\rho(t) a^{\dagger} a^{\dagger} aa$ is 
$2 \eta^2 \sigma^4 t^2$.  We will apply these results below to a discussion 
of the variance of $N$ at large times.  

The same strategy that we have just followed can be used to find a 
generating function for the expectations of general polynomials in 
the operators $q$ and $p$ in the free particle case.  Here the Hamiltonian   
is $H=p^2/(2m)$, and the equation to be solved is   
$$\eqalign{
{d \langle\langle B \rangle\rangle \over dt} =& 
{\rm Tr}\big( {d \rho \over dt} B+\rho {\partial B \over \partial t} \big)\cr
=&{\rm Tr}\Big( {\partial B \over \partial t}  
-{i\over \hbar} [B,H] -{1\over 2} \eta  
 [[B,q],q] \Big)  \rho~~~.\cr
}\eqno(17a)$$
We consider now the generating function defined by 
$$K_{\alpha \beta}^f={\rm Tr}B\rho(t)=  {\rm Tr}\big[ 
\exp\big(\alpha(q-tp/m)\big) \exp(\beta p) \rho(t)\big]~~~.\eqno(17b)$$
Using the fact that 
$$\exp\big(-(i/\hbar)t p^2/(2m)\big) q\exp\big((i/\hbar)t p^2/(2m)\big) 
=q-tp/m~~~,\eqno(17c)$$
we see that the terms $\partial B/\partial t$ and $-(i/\hbar)[B,H]$ 
in Eq.~(17a) cancel, so that we are left with 
$${d \over dt} K_{\alpha \beta}^f(t)={\rm Tr}\big(-{1\over 2}\eta
[[B,q],q]\rho(t)\big)~~~.\eqno(18a)$$
Using now the identity 
$$\exp\big(\alpha(q-tp/m)\big) \exp(\beta p) =
\exp(\alpha q)\exp\big(p(\beta-\alpha t/m)\big)
\exp\big(\alpha^2i\hbar t/(2m)\big)~~~,\eqno(18b)$$
the right hand side of Eq.~(18a) is easily evaluated to give 
$${1\over 2}\eta \hbar^2(\beta-\alpha t/m)^2 K_{\alpha \beta}^f(t)~~~.
\eqno(18c)$$
Equations (18a) and (18c) now give a differential equation that can be 
immediately integrated, giving a result analogous in form to Eq.~(16b), 
$$K_{\alpha \beta}^f(t)=
\exp\left[{1\over 6} \eta \hbar^2 t \big(3 \beta^2-3\beta\alpha t/m
+ \alpha^2 t^2/m^2 \big) \right]  K_{\alpha \beta}^f(0)~~~,
\eqno(19a)$$
with 
$$K_{\alpha \beta}^f(0)={\rm Tr}\big(\exp(\alpha q)\exp(\beta p) 
\rho(0) \big)
~~~.\eqno(19b)$$
This equation gives a generating function from which the results of 
Appendix E of Ghirardi, Rimini, and Weber [11] and their extensions to 
higher order polynomials, can be readily extracted.  In particular, 
expanding Eq.~(19a) through second order in $\alpha$ and $\beta$, one gets 
for the leading $\eta$ dependence of the 
expectations of quadratic polynomials  
in $p$ and $q$ the expressions given above in Eq.~(13c).   

Returning to the harmonic oscillator, the same methods 
can be applied to the generating function for general 
matrix elements of $\rho(t)$, although the results in this case are not so 
simple.  Let us define the generating function 
$$ 
L_{\alpha\beta}(t)={\rm Tr}\big( \exp(\alpha a^{\dagger}e^{-i\omega t}) 
|0\rangle \langle 0|
\exp(\beta a e^{i\omega t}) \rho(t)\big)~~~,\eqno(20a)$$
where $|0\rangle$ is the oscillator ground state obeying 
$a |0\rangle = \langle 0| a^{\dagger} =0$. (With the inclusion of a
normalization factor $\exp\big(-{1\over 2}(|\alpha|^2 + |\beta|^2)\big)$, 
this expression gives directly the matrix element 
of $\rho(t)$ between coherent states of the oscillator parameterized by 
$\alpha$ and $\beta$.)  When we take the time 
derivative of this expression, and apply Eq.~(14b), we now find that 
there are additional terms where an $a^{\dagger}$ multiplies $|0\rangle$ 
from the left, or an $a$ multiplies $\langle 0|$ from the right.  These 
can be converted to derivatives of $L_{\alpha \beta}$ with respect to the 
parameters $\alpha$ and $\beta$, and so we end up with the differential 
equation 
$$\eqalign{
{d \over dt} L_{\alpha \beta}(t) =&
\left\{ 
{i\over \hbar}  \left[
\left(\alpha -{\partial \over \partial \beta}\right)   
 e^{-i\omega t} \overline d(t)
- \left(\beta- {\partial \over \partial \alpha}\right)
e^{i\omega t}  d(t)  \right]\right.   \cr
-&{1\over 2} \eta \sigma^2\left. \left[
\left(\alpha - {\partial \over \partial \beta}\right)  e^{-i\omega t}  
 -  \left( \beta-{\partial \over \partial \alpha} \right)  e^{i\omega t}
\right]^2 \right\} L_{\alpha\beta}(t)~~~,\cr
}\eqno(20b)$$
which corresponds to making the substitutions $ \alpha \to \alpha -
{\partial \over \partial \beta}$, $\beta \to \beta -
 {\partial \over \partial \alpha}$ 
in Eq.~(15).  
Since the operators ${\partial \over \partial \beta} -\alpha$ 
and $ {\partial \over \partial \alpha} -\beta $ commute with 
one another, this equation  
can be formally integrated without requiring a time ordered product. Using   
$$\eqalign{
\left( {\partial \over \partial \alpha} -\beta \right) 
=& e^{\alpha \beta} {\partial \over \partial \alpha}e^{-\alpha \beta} ~~~,\cr
\left( {\partial \over \partial \beta} -\alpha \right) 
=& e^{\alpha \beta} {\partial \over \partial \beta}e^{-\alpha \beta} ~~~,\cr
}\eqno(21a)$$
the result can be compactly written as 
$$L_{\alpha \beta}= e^{\alpha \beta} 
\exp\left[\eta \sigma^2 t {\partial \over \partial \beta}
{\partial \over \partial \alpha} - {1\over 2} \eta \sigma^2  
{\sin \omega t \over \omega} \left( e^{-i\omega t} 
\left({\partial \over \partial \beta}\right)^2 + e^{i \omega t} 
\left({\partial \over \partial \alpha}\right)^2 \right) \right]
e^{-\alpha \beta} L^0_{\alpha \beta}(t)~~~,\eqno(21b)$$
with $L^0_{\alpha \beta}(t)$ the generating function in the absence of 
decoherence (that is, with $\eta=0$), which is given by 
$$L_{\alpha \beta}^0(t)=\exp\left( -{1 \over \hbar^2} |D(t)|^2 
+{i\over \hbar} [\overline D(t) \alpha - D(t) \beta ] \right)
{\rm Tr} e^{a^{\dagger}[\alpha + (i/\hbar) D(t)]} |0\rangle \langle 0|
e^{a[\beta -(i/\hbar) \overline D(t)]} \rho(0)~~~.\eqno(21c)$$
An alternative form of this result is obtained by introducing the 
{}Fourier transform of $e^{-\alpha\beta} L^0_{\alpha \beta}$ 
with respect to $\alpha$ and $\beta$, 
$$ e^{-\alpha \beta} L^0_{\alpha \beta}(t)=
\int dp_{\alpha} dp_{\beta} F(p_{\alpha},p_{\beta},t) 
e^{i\alpha p_{\alpha}+i\beta p_{\beta}}~~~,\eqno(22a)$$ 
in terms of which Eq.~(21b) takes the form
$$\eqalign{
L_{\alpha \beta}(t)=&e^{\alpha\beta} \int dp_{\alpha}dp_{\beta} 
\exp\left[-\eta \sigma^2 t   p_{\alpha} p_{\beta} 
+ {1\over 2} \eta \sigma^2  
{\sin \omega t \over \omega} \left( e^{-i\omega t} p_{\beta}^2 
+ e^{i \omega t}p_{\alpha}^2 \right) \right] \cr
\times&F(p_{\alpha},p_{\beta},t)   e^{i\alpha p_{\alpha}+i\beta p_{\beta}}
~~~.\cr
}\eqno(22b)$$
Thus, matrix elements of the density matrix for the decoherent forced 
oscillator can be explicitly (if formally) expressed in terms of 
matrix elements of the oscillator in the absence of decoherence.  

We have seen that exact results can be obtained for a number of 
properties of the density matrix evolution equation of Eq.~(5b).  This 
might have been suspected from the fact that earlier work [8] has shown  
that this equation leads to an exactly solvable expression for the fringe 
visibility in a proposed mirror superposition experiment described by   
an oscillator Hamiltonian.  More general 
density matrix evolution equations for a damped harmonic oscillator have 
been discussed in the literature [14].  When additional 
decoherence terms of the form $[c_1 q+ c_2 p,[c_1 q+ c_2 p,\rho]]$ 
(for general constants $c_{1,2}$) are added to the density matrix 
evolution equation for the forced oscillator, an explicit 
result for $K_{\alpha \beta}(t)$ 
generalizing Eq.~(16b) can still be easily obtained.  When dissipative terms  
proportional to a linear combination of 
$i[q,\{p,\rho\}]$   and  $i[p,\{q,\rho\}]$,   with 
$\{~,~\}$ the anticommutator, are added to the density matrix evolution 
equation, the differential equation for $K_{\alpha \beta}(t)$ contains 
terms involving $\partial/ \partial \alpha$ and $\partial/ \partial \beta$, 
and we then can no longer obtain an explicit formula for the expectation of 
the generating function analogous to Eq.~(16b).  Such dissipative  
terms are included in the evolution equations discussed in refs. [14], where    
some exact results are obtained.  We remark, however, 
that for mechanical or electrical systems with a very high 
quality factor $Q$, it 
can be a useful first approximation to neglect classical damping 
in studying stochastic reduction and decoherence effects, as done in the 
analysis of this paper.  [For the benefit of the reader familiar with the 
Lindblad form of the density matrix evolution equation, we give in Appendix 
B its relation to the commutator/anticommutator structures discussed here.]
\bigskip 
\centerline{\bf 6.~~ Bounds on  variances for unravelings}
\bigskip
So far we have studied quantum expectations of physical quantities in the 
mixed state density matrix $\rho$ obtained as the stochastic expectation of 
the pure state density matrix $\hat \rho$ that obeys Eq.~(5a).  In any given 
run of the physical process (or ``unraveling'' in the stochastics literature 
parlance), the quantum expectation of a physical quantity represented by 
a non-stochastic operator $B$ will be 
governed by ${\rm Tr} \hat \rho B$.  As before, let us use the notation 
$\langle 
\cdot\cdot\cdot \rangle$ to denote expectations formed with respect 
to $\hat \rho$, and the notation 
$\langle \langle \cdot\cdot\cdot \rangle\rangle$
to denote expectations formed with respect to $\rho=E[\hat \rho]$.  Then 
by linearity we evidently have 
$$\langle\langle B \rangle\rangle = E[\langle B \rangle]~~~.\eqno(23)$$
We shall now show that the variances corresponding to the single and double 
averages are related by an inequality.  Let 
$$\langle (\Delta B)^2 \rangle = {\rm Tr} \hat \rho (B-{\rm Tr}\hat \rho B)^2      
= {\rm Tr} \hat \rho B^2 - ({\rm Tr} \hat \rho B)^2~~~\eqno(24a)$$ 
be the squared variance of $B$ formed with respect to $\hat \rho$, and   
$$\langle\langle (\Delta B)^2 \rangle\rangle = 
{\rm Tr} \rho (B-{\rm Tr} \rho B)^2
= {\rm Tr} \rho B^2 - ({\rm Tr} \rho B)^2~~~\eqno(24b)$$ 
be the corresponding squared variance of $B$ formed with respect to $\rho$. 
The first of these two squared variances  
fluctuates from unraveling to unraveling; taking its expectation 
over the stochastic process we have 
$$\eqalign{
E[\langle (\Delta B)^2 \rangle]=& {\rm Tr} \rho B^2 
-E[ ({\rm Tr} \hat \rho B)^2  ]  \cr
=&{\rm Tr} \rho B^2 - ({\rm Tr} \rho B)^2   
+ C~~~,\cr
}\eqno(25a)$$
with $C$ a correction term given by 
$$\eqalign{
C=& ({\rm Tr}E[\hat \rho] B)^2 -  E[ ({\rm Tr} \hat \rho B)^2  ]   \cr
=&-E[({\rm Tr} \hat \rho B - {\rm Tr} E[\hat \rho] B)^2 ] \leq 0~~~.\cr
}\eqno(25b)$$
Hence we have obtained the inequality 
$$E[\langle (\Delta B)^2 \rangle]\leq {\rm Tr} \rho B^2 - ({\rm Tr} \rho B)^2 
=\langle \langle (\Delta B)^2 \rangle \rangle~~~,\eqno(26)$$
in other words, the squared variance formed from $\rho$ gives an upper 
bound to the expectation of the squared variance formed from $\hat \rho$.  
These results, and those of Sec. 3, can be used to calculate    
bounds on the expected variances $E[\langle (\Delta X_{1,2})^2 \rangle]$.  
When the effects of the driving terms $d(t),~\overline d(t)$ 
can be neglected (or at least remain bounded),
we see, for example, that at large times we have from Eq.~(10)
$$E[\langle (\Delta X_{1,2})^2 \rangle] \leq    
 {\rm Tr} \rho(t) X_{1,2}^2 -({\rm Tr} \rho(t) X_{1,2})^2
\simeq 2 \eta \sigma^4 t ~~~,\eqno(27a)$$
giving a large time bound on the mean squared stochastic 
fluctuations of $X_{1,2}$.  Similarly, we find \big(using the discussion 
following Eq.~(16d)\big) that when the effects of 
the driving terms can be neglected, the leading large time 
variance of $N$ is bounded by 
$$\eqalign{
E[\langle (\Delta N)^2 \rangle] \leq & {\rm Tr} \rho(t) N^2    
-({\rm Tr} \rho(t) N)^2 ={\rm Tr} \rho(t) (a^{\dagger}a^{\dagger}aa 
+a^{\dagger} a) - ({\rm Tr} \rho(t) a^{\dagger} a)^2  \cr
\simeq & 2\eta^2\sigma^4 t^2 - (\eta \sigma^2 t)^2 = 
\eta^2 \sigma^4 t^2~~~.\cr   
}\eqno(27b)$$    
Thus the root mean square variance in $N$, and the expectation of $N$, have 
the same time rate of growth.  
\bigskip
\centerline{\bf 7.~~Perturbation analysis for stochastic fluctuations}
\bigskip
We conclude our theoretical analysis by developing a 
formal perturbation theory for 
solving the evolution equation of Eq.~(5a) for the pure state density matrix 
$\hat \rho$.  Let $ \rho^{(0)}$ obey the evolution equation 
$$d  \rho^{(0)}= -{i\over \hbar}[H, \rho^{(0)}] dt,~~~\eqno(28a)$$
which holds when there are no stochastic terms, 
and let us expand the solution $\hat \rho$ of the corresponding stochastic 
equation as 
$$\hat \rho= \rho^{(0)} + \surd \eta \hat \rho^{(1/2)}
+ \eta \hat \rho^{(1)} +...~~~.\eqno(28b)$$
Inserting this expansion into Eq.~(5a), and equating like powers of  
$\eta$ on left and right, we get the following stochastic differential 
equations for $\hat \rho^{(1/2)}$ and $ \hat \rho^{(1)}$, 
$$\eqalign{
d \hat \rho^{(1/2)} =& -{i\over \hbar} [H,\hat \rho^{(1/2)}] dt 
+ \sigma [\rho^{(0)},[\rho^{(0)},a+a^{\dagger}]] dW_t~~~,\cr
d \hat \rho^{(1)}=&  -{i\over \hbar} [H,\hat \rho^{(1)}] dt
-{1\over 2} \sigma^2 [a+a^{\dagger},[a+a^{\dagger}, \rho^{(0)}]] dt \cr
+&\sigma\big(   [\rho^{(0)},[\hat \rho^{(1/2)},a+a^{\dagger}]]
+ [\hat \rho^{(1/2)},[\rho^{(0)},a+a^{\dagger}]] \big) dW_t~~~.\cr
}\eqno(28c)$$
The first step in solving these equations is to eliminate the time 
evolution associated with the Hamiltonian term $\hbar \omega a^{\dagger}a$ 
by defining, for any operator B, an interaction picture operator $B^I$ 
given by
$$B^I=e^{i\omega a^{\dagger}at} B e^{-i\omega a^{\dagger}at}~~~,\eqno(29a)$$
so that in particular 
$$a^I= a e^{-i \omega t}~~,~~~a^{I\dagger}=a^{\dagger} e^{i \omega t}~~~.
\eqno(29b)$$
Then in interaction picture, Eqs.~(28a)  and (28c) become 
$$\eqalign{
d\rho^{I(0)}= &-{i\over \hbar} [h^I,\rho^{I(0)}] dt~~~,\cr
d \hat \rho^{I(1/2)} =& -{i\over \hbar} [h^I,\hat \rho^{I(1/2)}] dt 
+ \sigma [\rho^{I(0)},[\rho^{I(0)},ae^{-i\omega t}+a^{\dagger}e^{i \omega t}]
] dW_t~~~,\cr
d \hat \rho^{I(1)}=&  -{i\over \hbar} [h^I,\hat \rho^{I(1)}] dt
-{1\over 2} \sigma^2 [ae^{-i\omega t}+a^{\dagger}e^{i\omega t},
[ae^{-i\omega t}+a^{\dagger}e^{i\omega t}, \rho^{I(0)}]] dt  \cr
+& \sigma\big(   [\rho^{I(0)},[\hat \rho^{I(1/2)},ae^{-i\omega t}
+a^{\dagger}e^{i \omega t}]]
+ [\hat \rho^{I(1/2)},[\rho^{I(0)},ae^{-i\omega t}
+a^{\dagger}e^{i\omega t}]] \big) dW_t~~~.\cr
}\eqno(29c)$$
Here $h^I=h^I(t) $ denotes the interaction picture form 
of the oscillator driving terms in the Hamiltonian, 
$$h^I(t)=d(t)a^{\dagger}e^{i \omega t}  + \overline d(t) a e^{-i\omega t}~~~.
\eqno(29d)$$
We can now deal with the $h^I$ term in the equations of motion by 
introducing an operator $U^I(t)$ that obeys the differential equation 
$${d U^I(t) \over dt} = -{i \over \hbar} h^I(t) U^I(t)~,~~
{dU^I(t)^{\dagger} \over dt} = {i \over \hbar} U^I(t)^{\dagger} h^I(t)~~~,
\eqno(30a)$$
which, using the definitions of Eq.~(16a),  
can be explicitly integrated to give 
$$\eqalign{
U^I(t)= &
\exp\left(-{i\over \hbar}\overline D(t)a \right) 
\exp\left(-{i\over \hbar}D(t) a^{\dagger}\right)
\exp\left({1\over \hbar^2}\int_0^t du  d(u) 
e^{i\omega u} \overline D(u)  \right)~~~. \cr
}\eqno(30b)$$
We can now use $U^I$ as an integrating factor to  integrate Eq.~(29c), 
giving finally explicit formulas for $ \rho^{I(0)}$, 
$\hat\rho^{I(1/2)}$, and $\hat \rho^{I(1)}$, 
$$\eqalign{
 \rho^{I(0)}(t)=&U^I(t) \rho^{I(0)}(0) U^I(t)^{\dagger} ~~~,\cr
\hat\rho^{I{1\over 2}}(t)= &\sigma \int_0^t U^I(t) U^I(s)^{\dagger} 
[ \rho^{I(0)}(s),[ \rho^{I(0)}(s),
ae^{-i\omega s}+a^{\dagger}e^{i\omega s}]] 
U^I(s) U^I(t)^{\dagger} dW_s~~~,\cr
\hat \rho^{I(1)}(t)=&-{1\over 2} \sigma^2 \int_0^t U^I(t) U^I(s)^{\dagger} 
[ae^{-i\omega s}+a^{\dagger}e^{i\omega s},[ae^{-i\omega s}
+a^{\dagger}e^{i\omega s}, \rho^{I(0)}(s)]]U^I(s)U^I(t)^{\dagger} ds\cr
+&\sigma \int_0^t U^I(t)U^I(s)^{\dagger}\Big( [ \rho^{I(0)}(s), 
 [\hat \rho^{I(1/2)}(s), ae^{-i\omega s}+a^{\dagger}e^{i\omega s} ]] \cr
 +&[\hat \rho^{I(1/2)}(s),[ \rho^{I(0)}(s),
 ae^{-i\omega s}+a^{\dagger}e^{i\omega s}]] \Big) U^I(s) U^I(t)^{\dagger} 
 dW_s~~~.\cr
 }\eqno(31a)$$
These equations give terms in the expansion in powers of $\surd\eta$ 
of the interaction picture quantity $\hat \rho^I$; to transform back 
to the original Schr\"odinger picture, one 
uses the inverse of Eq.~(29a), 
$$B=e^{-i\omega a^{\dagger}at} B^I e^{i\omega a^{\dagger}at}~~~,\eqno(31b)$$
taking $B$ to be successively $\hat \rho^{I(0,1/2,1)}$.  From the results  
of this calculation, one can in principle compute the quantum expectations 
${\rm Tr} \hat \rho B$  corresponding to different 
unravelings of the stochastic process. 

We see from Eq.~(31a) that the expression for $\hat \rho^{(1/2)}$ involves 
a stochastic integration over $dW_s$ with a non-stochastic integrand, and 
so as expected we have $E[\hat \rho^{(1/2)}(t)]=0$.  Additionally, we note 
that the expression for $\hat \rho^{(1)}$ contains two integrals, 
an ordinary integral involving an integration over $ds$, and a stochastic 
integral involving an integration over $dW_s$.  Since the integrand of the 
latter contains only stochastic quantities depending 
\big(through $\hat \rho^{I(1/2)}(s)$\big) on $dW_t$ for $t \leq s$, 
the stochastic  
expectation of the integral over $dW_s$ is zero, and so we have 
$$E[\hat \rho^{I(1)}(t])=-{1\over 2} \sigma^2 \int_0^t U^I(t) 
U^I(s)^{\dagger} 
[ae^{-i\omega s}+a^{\dagger}e^{i\omega s},[ae^{-i\omega s}
+a^{\dagger}e^{i\omega s}, \rho^{I(0)}(s)]]U^I(s)U^I(t)^{\dagger} ds
~,\eqno(32)$$
which writing $ \rho^{I(1)}(s)=  E[\hat\rho^{I(1)}(s)]$ gives the first 
term in the perturbation expansion for the ensemble expectation 
density matrix $\rho(t)$ obeying Eq.~(5b).  

Let us now use the results, $E[\hat \rho^{(1/2)}(t)]=0$ and
$E[\hat \rho^{(1)}(t)]=\rho^{(1)}(t) $ to interpret the inequality 
derived in Sec.~6.  Inserting the expansion for $\hat \rho$ into the 
definition of Eq.~(24a), we have 
$$\langle (\Delta B)^2 \rangle       
={\rm Tr}( \rho^{(0)}+\surd \eta \hat \rho^{(1/2)}+\eta \hat \rho^{(1)}) 
B^2 
- ({\rm Tr}( \rho^{(0)}+\surd \eta \hat \rho^{(1/2)}+\eta \hat \rho^{(1)})    
B)^2~~~.\eqno(33a)$$ 
Taking now the expectation of this equation, we get for non-stochastic 
operators $B$, 
$$E[\langle (\Delta B)^2 \rangle]  
={\rm Tr}(\rho^{(0)}+\eta  \rho^{(1)}) B^2 
- ({\rm Tr}( \rho^{(0)}+\eta \rho^{(1)}) B)^2
-\eta E[({\rm Tr}\hat \rho^{(1/2)} B)^2]  +{\rm O}(\eta^2)
~~~.\eqno(33b)$$ 
But comparing now with Eq.~(24b), we see that this is just 
$$E[\langle (\Delta B)^2 \rangle]  
= \langle\langle (\Delta B)^2 \rangle\rangle 
- \eta E[({\rm Tr}(\hat \rho^{(1/2)} B)^2]  +{\rm O}(\eta^2)
~~~,\eqno(33c)$$ 
in agreement with the expansion of the inequality of Eq.~(26) through 
terms of first order in $\eta$, and giving us insight into why the   
inequality takes this form.  By writing Eq.~(31a) for $\hat \rho^{(1/2)}$ 
in the form 
$$\hat \rho^{(1/2)}(t)= \int_0^t P(s,t) dW_s~~~,\eqno(34a)$$
with $P(s,t)$ denoting the integrand in Eq.~(31a), and using the It\^o 
isometry given in Appendix A, the stochastic expectation in the 
final term in Eq.~(33c) can be explicitly evaluated as an ordinary integral, 
$$ E[({\rm Tr}(\hat \rho^{(1/2)} B)^2]
= \int_0^t ds ({\rm Tr} P(s,t) B)^2~~~.\eqno(34b)$$
\bigskip
\centerline{\bf 8.~~ Estimates for  gravitational wave detection}
\centerline{\bf and nanomechanical oscillator experiments}  
\bigskip
Let us now use the results of the preceding  sections to make estimates 
for precision experiments involving monitoring of harmonically bound or 
free masses.  We begin by collecting the relevant formulas. For the harmonic  
oscillator, we have seen in Eq.~(9b) 
that the double expectation of the occupation 
number $N=a^{\dagger}a$ has a secular growth given by 
$$\langle\langle N \rangle\rangle \simeq \eta \sigma^2 t.~~~\eqno(35a)$$  
Since by our definition 
of Eq.~(2a), $\sigma =(\hbar/2m\omega)^{1\over 2}$, we have [10] 
$\sigma=\Delta X_{\rm SQL}$, with $\Delta X_{SQL}$ the so-called ``standard 
quantum limit'' for a conventional amplitude-and-phase measurement of 
$X_1$ or $X_2$, and so we can rewrite Eq.~(35a) as   
$$\langle\langle N \rangle\rangle\simeq \eta (\Delta X_{SQL})^2 t
~~~.\eqno(35b)$$
We have also seen in Eq.~(27b) that the right-hand side of Eq.~(35b) 
also gives at large times an upper 
bound to the root mean square variance in $N$, 
$$E[\langle(\Delta N)^2\rangle]^{1\over 2} \leq 
\eta (\Delta X_{SQL})^2 t~~~.\eqno(35c)$$
{}For the quantum nondemolition variables $X_{1,2}$, we have seen in 
Eq.~(10) that the double expectation is not influenced by stochastic 
reduction or decoherence effects, 
$$\delta \langle\langle X_{1,2} \rangle\rangle=0~~~,\eqno(35d)$$
while from Eq.~(27a) we get at large times an upper bound to the root 
mean square variances in $X_{1,2}$, 
$$E[\langle(\Delta X_{1,2})^2\rangle]^{1\over 2} 
\leq (2 \eta  t)^{1\over 2} \sigma^2
=(2 \eta t)^{1\over 2} (\Delta X_{SQL})^2~~~.\eqno(36a)$$
This last equation has a similar form to the corresponding equation for 
a free particle, for which the standard quantum limit $\Delta q_{SQL}$ 
in a position measurement is given [10] by
$$\Delta q_{SQL} = (\hbar t/m)^{1\over 2}~~~,\eqno(36b)$$
so that at large times we have from Eq.~(13c), 
and the fact that $\delta {\rm Tr}\rho q=0$, 
$$E[\langle(\Delta q)^2\rangle]^{1\over 2} 
\leq (\delta {\rm Tr} \rho q^2)^{1\over 2}
=(\eta t/3)^{1\over 2} \hbar t/m = (\eta t/3)^{1\over 2}
(\Delta q_{SQL})^2~~~.\eqno(36c)$$
These equations will form the basis for our analysis of experiments in 
which oscillating or free masses are monitored. Since we are making only  
order of magnitude estimates, we shall neglect numerical factors of order 
unity (such as the factor of 3 arising from generalizing from one to 
three dimensions), quoting all answers as powers of 10.  

To make estimates, we shall need values of both the stochasticity parameter 
$\eta$ and the elapsed time $t$.  
The value of $\eta$ depends on the stochastic 
reduction model under consideration.  In the GRW model [11] and also the 
QMUPL model [11], $\eta = \eta_0 N $, with $\eta_0 \sim 
10^{-2} {\rm s}^{-1} {\rm m}^{-2}$ and with $N$ the number of nucleons that   
are displaced in the measurement.  
{}For the 
CSL model, one has [8,11] $\eta=\gamma S^2 D^2 (\alpha/\pi)^{1\over 2}$, 
with $S$ the side length (for a cube of material), $D$ the density and 
$\gamma\sim  10^{-30} {\rm cm}^3 {\rm s}^{-1}$ and 
$\alpha \sim 10^{10} {\rm cm}^{-2}$ parameters of the model.  
We shall assume a nucleon density of $D \sim 10^{24} {\rm cm}^{-3}$,  and 
shall ignore the geometry dependence of $\eta$ by eliminating $S$ in terms of 
$D$ and the nucleon number $N$ by writing $S^2= (N/D)^{2\over 3}$, giving  
$$\eta =\gamma N^{2\over 3} D^{4\over 3}  (\alpha/\pi)^{1\over 2}~~~.
\eqno(37a)$$
{}For the elapsed time $t$ we shall take the inverse of the noise bandwidth 
frequency $F=\omega/(2Q)$, with $Q$ the quality factor, for the 
nanomechanical resonator experiment, and the inverse of the low frequency 
limit of the sensitive range for the gravitational wave detector experiments.  
Our reasoning here is that if accumulation of a small stochastic effect 
takes longer than the time estimated this way, the effects will be 
hard to distinguish from accumulated effects of the noise that sets the low  
frequency limit of the detector.  

The first experiment that we shall consider is the nanomechanical resonator 
reported by LaHaye et. al. [15], which uses a 19.7 MHz mechanical resonator 
containing $\sim 10^{12}$ nucleons, corresponding to 
$\Delta X_{SQL}\sim 10^{-14}$ m, and which has a noise bandwidth $F=903$ Hz.   
{}For the GRW and QMUPL models, we have 
$\eta \sim 10^{10} {\rm s}^{-1} {\rm m}^{-2}$, giving an accumulated 
$\langle\langle N \rangle\rangle$ in time $F^{-1}$ of $10^{-21}$, and a 
root mean square expected 
deviation in $X_{1,2}$ of $\sim 10^{-10} \Delta X_{SQL}$.  In the CSL 
model, $\eta \sim 10^{19} {\rm s}^{-1} {\rm m}^{-2}$, giving 
an accumulated $\langle\langle N\rangle\rangle$ in 
time $F^{-1}$ of $10^{-12}$, and 
a corresponding root mean square expected 
deviation in $X_{1,2}$ of $\sim 10^{-6} \Delta X_{SQL}$.  

The second experiment that we shall consider is the upgraded version of 
LIGO (the Advanced LIGO Interferometers), which monitors a quasi-free 
mass of 40 kg $\sim 10^{28}$ nucleons, 
and has a sensitive range extending down to  
$F \sim$ 70 Hz.  In the GRW model this apparatus has $\eta \sim 10^{27} 
{\rm s}^{-1} {\rm m}^{-2}$, while in the CSL model the value of this  
parameter is $\eta \sim 10^{30} {\rm s}^{-1} {\rm m}^{-2}$.  The standard 
quantum limit of Eq.~(36b) for position measurement over a 
time interval $t=(70 {\rm Hz})^{-1}$ is  $\Delta q_{SQL}\sim 
10^{-19}$ m, and we find that the root mean square stochastic deviation in  
the coordinate $q$ over this time interval is bounded 
by $\sim 10^{-7} \Delta q_{SQL}$ in 
the GRW model, and by  $\sim 10^{-5} \Delta q_{SQL}$ in the CSL model.  

The third experiment that we consider is the projected space-based 
Laser Interferometer Space Antenna (LISA) [17], which will monitor the 
positions of 
$2$ kg masses to an accuracy of $10^{-11}$ m, and which will be sensitive  
to frequencies down to $10^{-4}$ Hz. From Eq.~(36b), the standard  
quantum limit corresponding to a 2 kg  mass and $t \sim 10^4$ s is 
$10^{-15}$ m, in other words, this experiment will achieve a position 
accuracy of around $10^4$ times $\Delta q_{SQL}$.  For the CSL model, the 
corresponding root mean square stochastic deviation in the coordinate will 
be of order $100 \Delta q_{SQL}$, which is still a 
factor of 100 smaller than the observable displacement.

We see that in nanomechanical oscillator and Advanced LIGO experiments, 
predicted stochastic reduction effects 
are at least a factor of $10^{-5}$ below the relevant 
standard quantum limits,  and so are 
presently far from being detectable. The situation is better for  
LISA, where the stochastic reduction effect is predicted to be two orders of 
magnitude larger than the standard quantum limit, but still two orders of 
magnitude below the design position sensitivity.  Even though these 
experiments are not expected to observe an effect, they will place useful 
bounds on the stochasticity parameter $\eta$.  
Trying to do better will be a 
challenging goal for future experiments; clearly, the key will be achieving 
a much larger accumulation time $t$, corresponding to a greatly reduced 
noise bandwidth $F$ for the nanomechanical resonator, or a greatly reduced 
lower frequency limit for the gravitational wave detectors.  
We note in closing that when 
Eq.~(5b) is used as a model for environmentally induced (as opposed to 
postulated intrinsic) decoherence effects, the appropriate value of 
$\eta$ may be much larger than in the above estimates, and so 
in this case the 
effects for which we have obtained theoretical formulas may lie within reach 
of current experimental technique.

\bigskip
\centerline{\bf Acknowledgments}
\bigskip
This work was supported in part by the Department of Energy under
Grant \#DE--FG02--90ER40542, and I also acknowledge the hospitality of 
the Aspen Center for Physics, where this work was begun. I wish to thank   
Philip Pearle for checking Eq.~(13c) and reconciling it with the earlier 
calculation of GPR, Angelo Bassi and Emiliano Ippolito for 
checking the manuscript and a number of helpful suggestions, 
Kip Thorne for a helpful conversation and emails about LIGO, and 
Barry Barish for 
comments on future gravitational wave detectors.    
\bigskip
\centerline{\bf Appendix A: Basic It\^o Calculus Formulas}
\bigskip
The stochastic differential $dW_t$ behaves heuristically as a random 
square root of $dt$, as expressed in the It\^o calculus rules  
$$dW_t^2=dt~,~~~dW_t dt=dt^2=0~~~.\eqno(A1)$$
As a consequence of Eq.~(A1), the Leibniz chain rule of the usual calculus 
is modified to 
$$d(AB) = (dA)\; B + A \;dB + dA\; dB~~~. \eqno(A2)$$ 
Applying these two formulas to the definition 
$$\hat \rho(t)=|\psi_t \rangle \langle \psi_t|~~~,\eqno(A3)$$
the stochastic equation of motion of Eq.~(4b) for $|\psi_t\rangle$ is 
easily seen to imply the equation of motion of Eq.~(5a) for $\hat \rho(t)$.  
Because the It\^o differential $dW_t$ is statistically independent of 
the variables at and before time $t$, the final term in Eq.~(5a) vanishes 
when the stochastic expectation $E[\hat \rho]$ is taken, giving Eq.~(5b).  
Using this statistical independence of $dW_t$, and Eqs.~(A1)-(A3), 
we can get a useful 
formula for the expectation of a product of stochastic integrals.  
Consider the expectation  
$$f(t)= E[\int_0^t dW_u A(u) \int_0^tdW_u  B(u)]  ~~~,\eqno(A4)$$ 
which by Eq.~(A2) has the differential
$$ df(t) =   E[ dW_t A(t) \int_0^tdW_u  B(u) 
   +  \int_0^t dW_u A(u) dW_t  B(t) 
   + A(t) B(t) dt]= E[A(t) B(t)] dt~~~. \eqno(A5)$$ 
Integrating back using the right hand side of Eq.~(A5), we get  
$$ E[\int_0^t dW_u A(u) \int_0^tdW_u  B(u)]   =\int_0^t du E[A(u) B(u)] ~~~, 
\eqno(A6)$$
a formula called the It\^o isometry.  
\bigskip
\centerline{\bf Appendix B: Connection to the Lindblad Evolution Equation}
\bigskip
The most general completely positive density matrix evolution equation 
is given by the form studied by Lindblad [18] and Gorini, Kossakowski, and 
Sudarshan [18], generally referred to as the Lindblad equation, 
$${d\rho \over dt}=-{i\over \hbar} [H,\rho] 
+\sum_j(L_j \rho L_j^{\dagger} - {1\over 2} \{ L_j^{\dagger}L_j,\rho \})
~~~.\eqno(B1)$$
When $L_j$ is self-adjoint, so that $L_j=L_j^{\dagger}$, the 
summand in Eq.~(B1) reduces to the form 
$$-{1 \over 2} [L_j,[L_j ,\rho]]~~~,\eqno(B2)$$
which corresponds to the decoherence equation studied in the text when 
we take $L_j=q$, and more generally leads to a solvable oscillator model 
when $L_J=c_1q +c_2 p$ (with self-adjointness requiring real $c_{1,2}$.)
These two cases correspond respectively to repeated environmental (or 
intrinsic, in the case of reduction models) measurements of the system 
coupling to $q$ or to $c_1 q + c_2 p$.  

Dissipative equations in the Lindblad context are generated by taking $L_j$ 
to be non-self-adjoint.  For example, if we take $L_j=a$ for the harmonic 
oscillator, we get additional terms that cannot be represented as 
double commutators, as seen from the identity 
$$\eqalign{
4 \sigma^2 [a \rho a^{\dagger} -{1\over 2} \{a^{\dagger}a,\rho\}  ]
=&-{1\over 2} [q,[q,\rho]]-{1\over 2 m^2\omega^2}[p,[p,\rho]] \cr
-&{i\over 2 m\omega}([q,\{p,\rho\}]-[p,\{q,\rho\}])~~~,\cr
}\eqno(B3)$$
and a similar identity in which $a$ is interchanged with $a^{\dagger}$ 
and the $i$ on the right hand side is replaced by $-i$.  The paper of 
Salama and Gisin [14] studies a dissipative Lindblad 
equation with a term  
$L_j \propto a$, while the papers of Isar, Sandulescu, and Scheid [14] and 
Karrlein and Grabert [14] study a non-Lindblad master equation with 
a single dissipative term proportional to  $[q,\{p,\rho\}]$.  

\bigskip
\vfill\eject
\centerline{\bf References}
\bigskip
\noindent
[1]  J. J. Bollinger, D. J. Heinzen, W. M. Itano, S. L. Gilbert, and 
D. J. Wineland, Atomic physics tests of nonlinear quantum mechanics, in 
{\it Atomic Physics 12:  Proceedings of the 12th International Conference 
on Atomic Physics}, J. C. Zorn and R. R. Lewis, eds., American Institute 
of Physics Press, New York (1991).  \hfill\break
\bigskip 
\noindent
[2] I. Bialynicki-Birula and J. Mycielski, Nonlinear wave mechanics.  {\it 
Ann. Phys.} {\bf 100}, 62-93 (1976); S. Weinberg, Particle states as 
realizations (linear or nonlinear) of spacetime symmetries. {\it Nucl. Phys.   
B Proc. Suppl.} {\bf 6}, 67-75 (1989), {\it Spacetime Symmetries}, 
Y. S. Kim and W. W. Zachary, eds., North-Holland, Amsterdam; S. Weinberg, 
Precision tests of quantum mechanics, {\it Phys. Rev. Lett.} {\bf 62}, 
485-488 (1989); S. Weinberg, Testing quantum mechanics, {\it Ann. Phys.} 
{\bf 194}, 336-386 (1989).  \hfill\break
\bigskip
\noindent
[3]  N. Gisin, Stochastic quantum 
dynamics and relativity, {\it Helv. Phys. Acta} 
{\bf 62}, 363-371 (1989); N. Gisin, Weinberg's non-linear quantum mechanics 
and supraluminal communications, {\it Phys. Lett. A} {\bf 143}, 1-2 (1990); 
J. Polchinski, Weinberg's nonlinear quantum mechanics and the Einstein-
Podolsky-Rosen paradox, {\it Phys. Rev. Lett.} {\bf 66}, 397-400 (1991); 
N. Gisin and M. Rigo, Relevant and irrelevant nonlinear Schr\"odinger 
equations, {\it J. Phys. A:  Math. Gen.} {\bf 28}, 7375-7390 (1995).  
\hfill\break
\bigskip
\noindent
[4] For reviews, see A. Bassi and G. C. Ghirardi, Dynamical reduction 
models, {\it Phys. Reports} 
{\bf 379}, 257-426 (2003); P Pearle, Collapse models, in {\it Open Systems 
and Measurements in Relativistic Quantum Field Theory}, Lecture Notes in 
Physics 526, H.-P. Breuer and F. Pettrucione, eds., Springer-Verlag, 
Berlin (1999).  See also S. L. Adler, {\it Quantum Theory as an Emergent 
Phenomenon}, Cambridge University Press, Cambridge, UK (2004), Chapt. 6.    
\hfill\break
\bigskip
\noindent
[5] For a brief catalog of the different forms of localization models,  
and references, see reference [6] of A. Bassi, E. Ippoliti, and S. L. Adler, 
Towards quantum superpositions of a mirror: stochastic collapse analysis, 
arXiv: quant-ph/0406108 (2004).  See also ref. [11] below.  
\bigskip
\noindent
[6]  S. L. Adler, ref. [4], p. 188.\hfill\break
\bigskip
\noindent
[7]  W. Sh\"ollkopf and J. P. Toennies, Nondestructive mass selection of 
small van der Waals clusters, {\it Science} {\bf 266}, 1345-1348     
(1994); M. Arndt, O. Nairz, J. Vos-Andreae, C. Keller, G. van der Zouw, 
and A. Zeilinger, Wave-particle duality of $C_{60}$ molecules, 
{\it Nature} {\bf 401}, 680-682 (1999); O. Nairz, M. Arndt, and A. Zeilinger, 
Experimental challenges in fullerene interferometry, {\it J. Mod. Optics} 
{\bf 47}, 2811-2821 (2000); O. Nairz, B. Brezger, M. Arndt, and A. Zeilinger, 
Diffraction of complex molecules by structures made of light, 
{\it. Phys. Rev. Lett.} {\bf 87}, 160401 (2001).  \hfill\break
\bigskip
\noindent
[8]  A. Bassi, E. Ippoliti, and S. L. Adler, ref. [5]; S. L. Adler, 
A. Bassi, and E. Ippoliti, Towards quantum superpositions of a mirror: 
stochastic collapse analysis -- calculational details, arXiv: 
quant-ph/0407084 (2004).\hfill\break
\bigskip
\noindent
[9] W. Marshall, C. Simon, R. Penrose, and D. Bouwmeester, Towards quantum 
superpositions of a mirror, {\it Phys. Rev. Lett.} {\bf 91}, 130401 (2003).
\hfill\break
\bigskip
\noindent
[10]  V. B. Braginsky and F. Ya. Khalili, {\it Quantum Measurements}, 
Cambridge University Press, Cambridge, UK (1991); C. M. Caves, K. S. Thorne, 
R. W. P. Drever, V. D. Sandberg, and M. Zimmerman, On the measurement of a 
weak classical force coupled to a quantum-mechanical oscillator. I. Issues 
of principle, {\it Rev. Mod. Phys.} {\bf 52}, 341-392 (1980); M. F. Bocko 
and R. Onofrio, On the measurement of a weak classical force coupled to 
a harmonic oscillator: experimental progress, {\it Rev. Mod. Phys.} {\bf 68}, 
755-799 (1996). \hfill\break
\bigskip
\noindent
[11]  The models for localization that are considered here are:  G. C. 
Ghirardi, A. Rimini, and T. Weber,  Unified dynamics for microscopic and 
macroscopic systems, {\it Phys. Rev. D} {\bf 34}, 470-491 (1986), known 
as GRW or as QMSL (Quantum Mechanics with Spontaneous Localizations); 
P. Pearle, Combining stochastic dynamical state-vector reduction with 
spontaneous localization, {\it Phys. Rev. A} {\bf 39}, 2277-2289 (1989),  
and G. C. Ghirardi, P. Pearle, and A. Rimini, Markov processes in Hilbert 
space and continuous spontaneous localization of systems of identical 
particles, {\it Phys. Rev. A} {\bf 42}, 78-89 (1990), known as CSL 
(Continuous Spontaneous Localization);  L. Di\'osi, Models for universal 
reduction of macroscopic quantum fluctuations, {\it Phys. Rev. A} {\bf 40}, 
1165-1174 (1989), known as QMUPL (Quantum Mechanics with Universal Position 
Localization).  
\bigskip
\noindent
[12] See, e.g., E. Joos and H. D. Zeh, The emergence of classical properties  
through interaction with the environment, {\it Z. Phys. B} {\bf 59}, 223-243  
(1985), Eq.~(3.59).\hfill\break
\bigskip
\noindent
[13] G. C. Ghirardi, P. Pearle, and A. Rimini, ref. [11], Eqs.~(3.38a-c) 
and (3.41a-c).\hfill\break  
\bigskip
\noindent
[14] G. S. Agarwal, Master equations in phase-space formulation of quantum 
optics, {\it Phys. Rev.} {\bf 178}, 2025-2035 (1969); G. S. 
Agarwal, Brownian motion of a quantum oscillator, 
{\it Phys. Rev. A} {\bf 4}, 739-747 (1971).  
Extensive further references  
can be found in: R. Karrlein and H. Grabert, 
Exact time evolution and master equations for the damped harmonic oscillator, 
{\it Phys. Rev. E} {\bf 55}, 153-164 (1997); A. Isar, A. Sandulescu, and 
W. Scheid, Purity and decoherence in the theory of a damped harmonic 
oscillator, {\it Phys. Rev. E} {\bf 60}, 6371-6381 (1999). \hfill\break
\bigskip
\noindent
[15]  M. D. LaHaye, O. Buu, B. Camarota, and K. C. Schwab, Approaching the   
quantum limit of a nanomechanical resonator, {\it Science} {\bf 304}, 
74-77 (2004).\hfill\break  
\bigskip
\noindent
[16]  A. Abramovici et. al., LIGO: The laser interferometer gravitational-
wave observatory, {\it Science} {\bf 256}, 325-332 (1992); B. C. Barish 
and R. Weiss, LIGO and the detection of gravitational waves, 
{\it Physics Today} {\bf 52}, 44-50 (1999); P. S. Shawhan, Gravitational 
waves and the effort to detect them, {\it American Scientist} {\bf 92}, 
350-357 (2004). \hfill\break  
\bigskip
\noindent
[17]  J. Alberto Lobo, LISA, ArXiv: gr-qc/0404079; R. Irion, Gravitational 
Wave Hunters Take Aim at the Sky, {\it Science} {\bf 297}, 1113-1115 (2002).
\hfill\break
\bigskip
\noindent
[18]  G. Lindblad, on the generators of quantum dynamical semigroups, 
{\it Commum. Math. Phys.} {\bf 48}, 119-130 (1976); V. Gorini, 
A. Kossakowski, and E. C. G. Sudarshan, Completely positive dynamical 
semigroups of $N$-level systems, {\it J. Math. Phys.} {\bf 17}, 
821-825 (1976). \hfill\break
\bigskip
\bye